\documentclass[prl,reprint,amsmath,amssymb,superscriptaddress,twocolumns]{revtex4-1}

\usepackage[utf8]{inputenc}
\usepackage{graphicx}
\usepackage{natbib}
\usepackage{gensymb}
\usepackage{xcolor}
\usepackage{cancel}
\usepackage{amsmath}
\usepackage{bm}
\usepackage{array}
\usepackage{amsfonts}
\usepackage{amssymb}
\usepackage{epstopdf}
\usepackage{subfiles}
\usepackage{soul}
\usepackage{float}

\newcommand{\dd}{\textrm{d}}
\newcommand{\Lk}{L_{\kappa}}
\newcommand{\lD}{\ell_{D}}

\newcommand{\bapt}[1]{\textcolor[RGB]{52, 103, 133}{#1}}

\begin{document}

\title{Wave-induced fracture of a sea ice analog}

\author{B. Auvity}
\author{L. Duchemin}
\author{A. Eddi}
\author{S. Perrard}

\affiliation{
PMMH Lab, ESPCI, CNRS, PSL University, Sorbonne Universit\'e, Universit\'e Paris Cit\'e\\
7 quai Saint Bernard, 75005 Paris, France
}

\begin{abstract}
We study at the laboratory scale the rupture of thin floating sheets made of a brittle material under a wave-induced mechanical forcing. We show that the rupture occurs where the curvature is maximum and the break-up threshold strongly depends on the wave properties. We observe that the critical stress for fracture depends on the forcing wavelength. Hence our observations are incompatible with a critical stress criterion for fracture. Instead, our measurements can be rationalized as an energy criterion: a fracture propagates when the material surface energy is lower than the released elastic energy, which depends on the forcing geometry. In light of these findings, it may be worthwhile to revisit current numerical models of sea ice fracture by ocean waves.
\end{abstract}

\maketitle

\textit{Introduction} - The freezing of ocean water forms sea ice, that covers the Arctic region and surrounds Antarctica. When exposed to ocean waves, sea ice breaks~\cite{Lu2008AerialOO,Asplin_2012,Liu1988WavePI} forming fragments of multiple sizes~\cite{Toyota_2006}. This phenomenon occurs in partially ice-covered region, called Marginal Ice Zone (MIZ)~\cite{Squire_2019,Dumont2022}. The MIZ plays a crucial role in the climate dynamics~\cite{Eayrs2019} and marine biology~\cite{Ardyna2014} of these regions. \bapt{Sea ice forms a thin brittle layer, and its fracture by waves is currently modelled by a critical strain~\cite{Voermans2020}}.

In general, the fracture of a thin sheet of brittle material depends on the forcing geometry and the presence of existing cracks~\cite{Dirgantara_2002,Hasebe1980}. The forcing geometry here results from the surface wave propagation under the ice. Therefore, the fracture threshold should result from an interplay between the mechanical properties of the sheet and the characteristics of the waves. 
In the absence of fracture, the deformation of a thin ($kh \ll 1$ with $k$ the wavenumber and $h$ the thickness), floating elastic sheet by surface waves has been previously studied, both in the linear~\cite{Domino_2018} or weakly non-linear regime~\cite{Deike_PRF_2017}. In the linear regime, neglecting the sheet inertia and surface tension, the deformation follows the dispersion relation of hydro-elastic waves~\cite{lamb1917},
\begin{equation}
\omega^2 = g k + \frac{D}{\rho} k^5,
\label{eq:HEW}
\end{equation}
where $\omega$ is the angular frequency, $k$ the wavenumber, $g$ the acceleration of gravity, $\rho$ the liquid density and $D$ the flexural modulus. The modulus $D$ is related to the mechanical properties of the sheet, $D = E h^3/(12(1-\nu^2))$, with $E$ the Young's modulus and $\nu$, Poisson's ratio. The elasto-gravity length $\lD = (D/\rho g)^{1/4}$ separates the gravity dominated regime ($k \lD \ll 1$) from the elastic regime $k \lD \gg 1$. For an ice fragment excited by free surface waves, the wavelength depends on the regime of wave propagation~\cite{Montiel_2013_1,Montiel_2013_2}, hence we expect the fracture threshold to depend on the ratio $k \lD$. Although the mechanical properties 
of sea ice are known to be highly variable, due to variations of composition, age or salt content~\cite{Petrovic_2003,timco2010}, the fracture criterion currently used in regional and global models is based on a critical strain~\cite{Voermans2020,Li_JMSE_2021}. This criteria depends neither on the ice mechanical properties nor on the flexural length $\lD$.

Previous works focusing on the wave fracturing of either natural sea ice at the geophysical scale~\cite{Dumas2023} or artificially grown ice at the laboratory scale~\cite{Herman_2019} have suggested that sea ice breaks where the curvature is maximum. However, the forcing geometry in these studies has not been systematically varied. At the laboratory scale, L. Saddier~\textit{et al.}~\cite{saddier2023breaking} recently studied the fracturing of a mimetic material. They use a granular raft formed by the capillary aggregation of micrometric particles~\cite{Planchette_2012} to investigate the fracture induced by waves. However, the raft broke under viscous stresses, a different fracture mechanism than bending induced fracture. In this letter, we study the fracture of a thin brittle sheet at the laboratory scale. The floating crust is made of a material brittle enough to break when bent by surface waves. Using stationary waves of varying wavelength, we discriminate between possible fracture criterion. We find that the critical stress depends on the forcing geometry, and we model the fracture threshold with an energetic criteria involving the sheet geometry. We eventually discuss the potential implication of our findings to the fracture of sea ice by ocean waves.

\textit{Experimental set-up -} We form a thin brittle solid crust floating on the water surface, using a commercial glue varnish (Vernis-colle spray Megacrea DIY 28715). The varnish is sprayed from a height of about $15~$cm on a water volume of $80$ x $28$ x $11~$cm$^3$. The varnish drops dry on the water surface, forming a floating, textured and semi-transparent solid crust. We combine pycnometer and weight measurements of the deposited mass to deduce the varnish thickness $h$ and density $\rho_v$. We find average values of $\rho_v = 680 \pm 26$ kg.m$^{-3}$ and $\Bar{h} = m_v/(\rho_v S_v) \simeq 100\mu$m, where $S_v$ is the covered area. For each varnish crust we observe spatial variations of its transparency. Using the Beer-Lambert absorption law, we infer a spatial thickness variation of 14\%. From one realisation to another, we also observe significant variations of the averaged thickness estimate, ranging from $57$ to $207\mu$m. As the varnish crust is composed of micro-droplets that solidify into coagulating grains, it also presents heterogeneities at the droplet scale, that is, on the order of tens of microns.

\begin{figure}[t!]
    \centering
\includegraphics[width = \columnwidth]{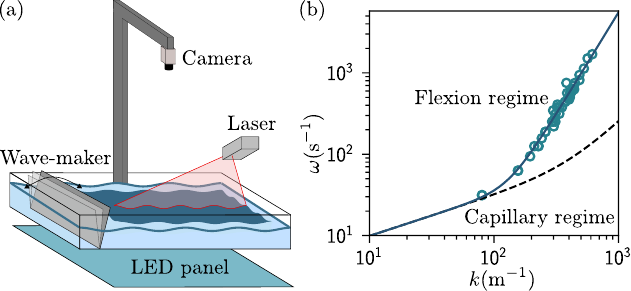}
    \caption{\label{fig:1}a) Sketch of the experimental set-up, showing the varnish layer and the one dimensional profilometry system. b) Experimental measurements of the dispersion relation of surface waves propagating under a varnish layer of thickness $h = 158 \pm 22~\mu$m (blue circles) obtained from a single experiment. 
   The gravito-capillary dispersion relation (black dashed line)is superimposed. The experimental data points lie on the flexural branch, $\omega^2 \sim k^5$.
   }
\end{figure}

To characterize the varnish mechanical properties, we use the experimental set up sketched in [Fig. \ref{fig:1}a)]. Sinusoidal plane waves are generated with a linear motor (LinMot DM01-23x80F-HP-R-60\_MS11), which actuates an inclined plate oriented with an angle -$10 ^\circ$ with the vertical. We vary concomitantly the wave frequency $f$ from $1$ to $100~$Hz and the amplitude $A$ from $10$ to $1$mm. For these small wave amplitudes, the varnish does not break, and waves propagate at the surface.

We use a one-dimensional profilometry system to extract the wave properties ($f$, $A$, wavenumber $k$). A laser sheet is projected onto the varnish surface with an angle $\theta$ ranging from $5$ to $30 ^\circ$ to adapt the magnification ratio, which varies between 1.7 and 11 depending on the required vertical resolution. Since the laser sheet is not significantly reflected by the water surface, all wave profiles are measured at the surface of the continuous varnish. We record top-view images of the varnish and the laser line with a Basler a2A1920-160ucBAS color camera of resolution $1920$ x $1200$ pixels, corresponding to a $48$ x $30$ cm field of view. The sampling frequency is chosen to capture a minimum of 5 images per wave period.
Using image processing, we extract the position of the laser line with sub-pixel accuracy. We obtain the wavenumber $k$ and the amplitude $A$ of the surface wave from spatial Fourier analysis. 
[Fig. \ref{fig:1}b)] shows the dispersion relation (blue diamond) 
of the surface waves, namely the angular frequency $\omega$ as a function of the wavenumber $k$ for a varnish layer of thickness $h = 158~\pm22\mu$m.  
The waves follow the dispersion relation (eq.~\ref{eq:HEW}). The flexural modulus $D$ depends on the Young's modulus $E$, the thickness $h$, and the Poisson's ratio $\nu$. Using $D$ as a fit parameter, we obtain a quantitative agreement with $D= 3.1 \pm 0.1$ $10^{-5}$ J.m$^{-2}$. We find that the varnish layer behaves as a solid. We associate a flexural length 
\begin{equation}
\lD = \left(\frac{D}{\rho g}\right)^{1/4} = 7.5 \pm 0.1~\textrm{mm},
\end{equation}
that separates the gravito-capillary wave regime ($k \lD < 1$) from the hydro-elastic ($k \lD > 1$) wave regime. Repeating the experiment for different varnish layers, we systematically extract the flexural modulus and the varnish thickness. We find an average varnish Young's modulus $E = 65 \pm 14$~MPa using $\nu = 0.4$.

The corresponding flexural length being always larger than the capillary length ($\ell_c \sim 2.4$~mm), capillarity is negligible for all wavelengths.

\begin{figure}[t]
    \centering
    \includegraphics[width = \columnwidth]{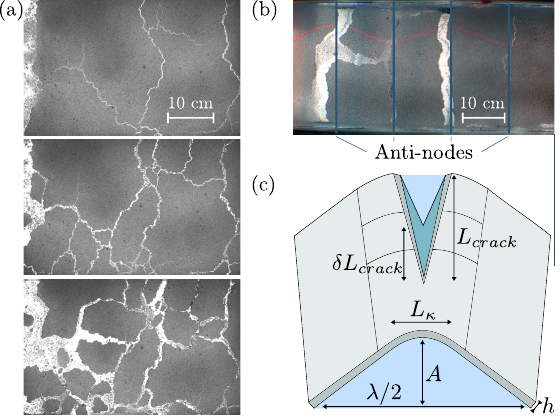}
    \caption{\label{fig:2} a) 3 Photographies issued from the fracture of a single varnish sheet ($h \simeq 150~\mu$m) by propagative waves ($f = 3.33~$Hz, $A = 6.7~$mm and $\lambda = 14~$cm), after 12, 105 and 203 periods. 2)b) Photography of a varnish layer after $307$ periods with $\lambda = 27~$cm, $A = 5.9~$mm, $f=2.397~$Hz and $h = 90 \pm 13~\mu$m. The varnish has broken on the anti-nodes of the waves. c) Schematical representation of the sheet geometry near a fracture tip. A planar, non-linear wave 
    locally induces a curvature on a characteristic length $L_{\kappa}$ in the transverse direction. For a large enough amplitude, a fracture of length $L_{crack}$ propagates along the wave crest/trough.}
\end{figure}

\textit{Fracturing with waves} - For larger wave amplitudes, the varnish layer fractures into multiple fragments. As an example, [Fig. \ref{fig:2}a)] shows 3 snapshots of a varnish layer seen from above, after respectively 12, 105 and 203 periods of oscillation, for $f = 3.33~$Hz, $A = 6.7~$mm, $h \simeq 150~\mu$m and a wavelength $\lambda = 14~$cm. 
We observe a complex dynamics of fragmentation: the orientation and the length of the first fractures present a wide distribution. On a longer time scale, the fragments collide and erode. This leads to the creation of very small fragments with a typical grain size of $100~\mu$m, while larger fragments become more circular.
At last, fragmentation and erosion continue until the layer reduces to grain-sized fragments.
To better control the geometry of the first fractures, we modify the setup to produce stationary, planar waves. For each forcing frequency, we adapt the length of the water tank to a multiple of half the wavelength. Under a stationary wave pattern, the stress field is such that the nodes maximize tension, whereas the crest/trough (anti-nodes) maximizes bending and the viscous stress applied by the liquid on the sheet bottom.
[Fig. \ref{fig:2}b)] presents a photograph of an experiment with 3 wavelengths after 307 forcing periods with $\lambda = 27~$cm, $A = 5.9~$mm, $f=2.397~$Hz and $h = 90 \pm 13~\mu$m and shows that the waves are non-linear, since $Ak = 0.14$.

\begin{figure}[t!]
    \centering
    \includegraphics[width =\columnwidth]{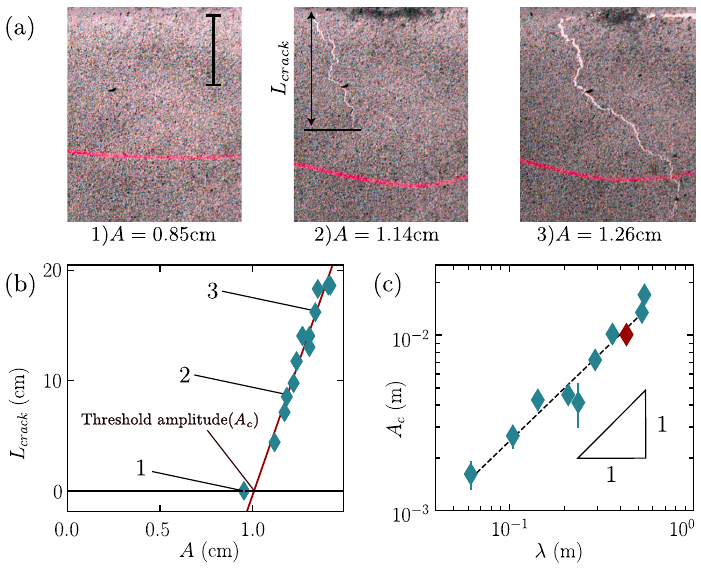}
    \caption{\label{fig:3} a) Crack propagation in a varnish layer of thickness $h = 57\pm8~\mu$m for waves ($\lambda=43.12~$cm) of increasing amplitude (images 1,2 and 3), respectively $A=0.96/1.18/1.34~$cm. b) Crack length $L_{crack}$ as a function of the wave amplitude $A$ corresponding to the experiment of [Fig. \ref{fig:3}a)]. A linear fit of the experimental data (red line) gives the threshold amplitude for break-up $A_c = 1.01~$cm corresponding to $L_{crack}=0$. 
    c) Wave amplitude threshold $A_c$ as a function of the wavelength $\lambda$ for different varnish layers, showing a linear relationship in a log-log scale. The varnish layer breaks in the non-linear wave regime of propagation, since $A_c k \simeq 0.16$. }
\end{figure}

The laser line used for the profilometry appears in red. Superimposed in solid blue is the position of the anti-nodes.
We observe that the varnish layer breaks along 4 lines parallel to the wave front separated by $\lambda / 2$. Occasionally, secondary fractures appear in the direction perpendicular to the wave front. For each experiment we performed, the primary fractures appear at the anti-nodes, which correspond to maxima of curvature $\kappa$, \textit{i.e.} loci of maximum flexural stress. The viscous stress is also maximum on the anti nodes. However, it decreases with the wavelength, and, as shown later in [Fig. \ref{fig:3}c)], our experimental results show that the fracture threshold in amplitude increases with the wavelength. Consequently, we rule out viscous stress as fracture mechanism. We thus conclude that the physical mechanism responsible for the first fracture of the varnish is bending rather than uniaxial tension or viscous stress. 


\textit{Fracture threshold} - 
To study the fracture threshold, we excite stationary waves starting from a water surface at rest. Doing so, we observe a peculiar transient regime. The wave amplitude grows to a maximum over the first few periods, then decay to approximately 70\% of the maximum to reach a stationary regime. As a consequence, crack evolution is observed only during the 2 or 3 periods over which the amplitude is near its maximum.

[Fig. \ref{fig:2}c)] presents a sketch of the fracture geometry in the vicinity of an existing crack. We assume that the waves are planar, with an amplitude $A$, a frequency $f$, and a wavelength $\lambda$. We define $\Lk$, the typical size on which the curvature is maximal, and $L_{crack}$, the length of the first crack. 

[Fig. \ref{fig:3}a)] shows an experiment with $f = 1.832~$Hz, $\lambda = 43.12~$cm and $h = 57 \pm 8~\mu$m for three increasing forcing amplitudes $A$. We observe that the varnish layer breaks above a critical threshold $A_c$, on a finite crack length $L_{crack}$ (images 1 to 3), which increases rapidly with increasing forcing amplitude.
To extract the threshold amplitude $A_c$ of crack apparition, we plot the crack length $L_{crack}$ as a function of $A$ on [Fig. \ref{fig:3}b)]. We observe that $L_{crack}(A)$ is an affine function of $A$ for $A>A_c$. 

We then vary the wave frequency and we measure the critical wave amplitude $A_c$ for a wavelength $\lambda$ ranging from $5$ to $50~$cm. Each fracture experiments is performed in the gravity regime since $k\ell_D < 1$. [Fig. \ref{fig:3}c)] shows that the threshold amplitude is proportional to the wavelength and corresponds to a constant steepness of $A_c k = 0.16$. 

\begin{figure*}[!t]
    \centering
    \includegraphics[width = \textwidth]{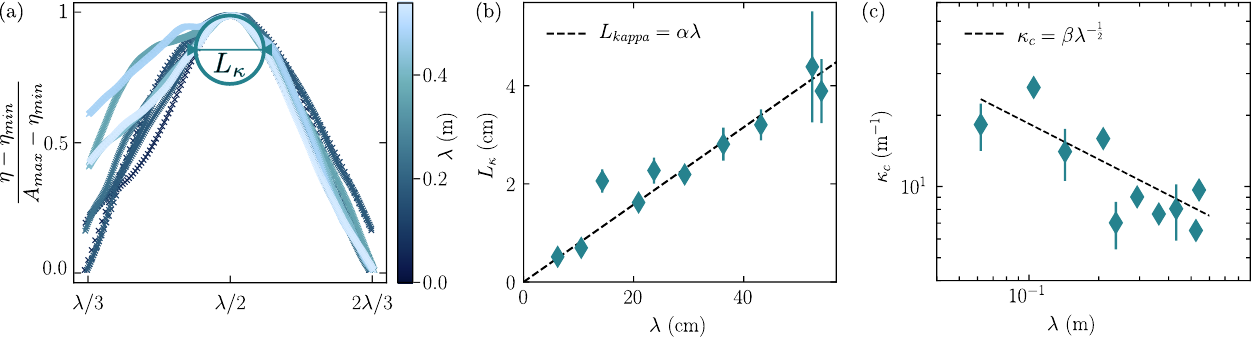}
    \caption{\label{fig:4}a) Wave profiles near the maximum of curvature for different wavelength. The  normalized  relative  elevation $\eta-\eta_{min}/(A_{max}-\eta_{min})$ is plotted as a function of the position $x$  in  unit  of $\lambda$. The wavelength varies from $6.2$ to $54.1$cm. The average $\Lk$, the half width at maximum curvature, is showed with a blue circle, highlighting the auto-similarity of the wave profiles as well as their non linearity.
    b) Half width of max curvature $L_{\kappa}$ as a function of the wavelength $\lambda$ for the closest wave profile to the threshold. It shows a linear relationship $\Lk = \alpha \lambda$, with $\alpha = 0.079\pm 0.004$.
    c) Curvature threshold $\kappa_c$ for fracture as a function of the wavelength $\lambda$, compatible with $\kappa_c = \beta \lambda^{-1/2}$ with $\beta = 5.82\pm0.6$ m$^{-1/2}$.
    }
\end{figure*}

However, the relevant quantity for fracture propagation induced by a bending load is the curvature $\kappa$ of the plate. For $A_c k \sim 0.16$, the wave profile is expected to be non linear, and we therefore investigate further the wave geometry near the maximum of curvature.
[Fig. \ref{fig:4}a)] presents the zoomed in wave profiles near the maximum of curvature, at the critical amplitude, for various wavelength. The normalized relative elevation $\eta-\eta_{min}/(A_{max}-\eta_{min})$ is plotted as a function of the position $x$ in unit of $\lambda$. The colormap represents the wavelength, from $6.2$ cm in dark blue to $54.1$ cm in light cyan. We observe that the wave profiles near their maximum are indeed non linear. The length $L_{\kappa}$ is defined as the half width at maximum curvature, and its average is represented with a blue circle. We observe that $L_{\kappa}$ is constant in unit of $\lambda$, showing graphically that $L_{\kappa}$ is proportional to the wavelength.

[Fig. \ref{fig:4}b)] shows $\Lk$ at threshold as a function of $\lambda$. The black dash line corresponds to a linear fit $\Lk = \alpha \lambda $ with $\alpha = 0.079 \pm 0.004$. In comparison, we expect $\alpha = 1/3$ for linear waves. We also measure the magnitude of the curvature peak $\kappa$, by fitting a second order polynomial on a length $\Lk$ around the maximum of amplitude. We assume that for weak nonlinear wave profiles, the curvature can be expressed as $\kappa = aA + bA^{2}$, with two unknown coefficients $a$ and $b$. We fit these coefficients to compute the curvature threshold $\kappa_c$ corresponding to the amplitude threshold $A_c$ for each wavelength.

[Fig. \ref{fig:4}c)] shows the critical curvature $\kappa_c$ as a function of the wavelength $\lambda$. We observe that the curvature threshold decreases significantly with the wavelength. Our observations \bapt{on the break-up of the varnish layer by waves} are therefore incompatible with a fracture criterion based on a critical stress $\sigma_c = E \kappa_c h$, independent of the wavelength.

\textit{Fracture models} - We propose an energy based criterion to model the crack propagation induced by the waves. Starting from an existing crack, such an energy criterion for fracture propagation results from the balance between the increment of surface energy $G$ required to form a new surface and the strain energy $\dd U_T$ released when a crack propagates on a surface $\dd S$. $G$ is also called energy release rate and is written~\cite{tada2000, anderson2005}:
\begin{equation}
G = -\frac{d U_T}{d S},
\label{eq:G}
\end{equation}
with $\dd S = 2 h \delta L_{crack}$ the crack surface. The elastic energy released by the crack propagation can be written as~\cite{timoshenko1959}:
\begin{equation}
U_T= D \iint_{\Sigma}\kappa^{2}(x,y) \,dx \,dy.
\label{eq:Ut}
\end{equation}
Where $\kappa$ is the curvature field on the sheet. For an imposed curvature, the elastic energy must be computed on the entire sheet.
In our situation, due to the imposed displacement by non linear water waves, the plate is only curved on a characteristic length $\Lk$. Once a crack of length $L_{crack} > \Lk$ is present, we assume that the elastic energy $U_T$ can only be released on the characteristic length $\Lk$ in the direction perpendicular to the crack, as sketched on [Fig. \ref{fig:2}c)]. 
The elastic energy is released on a surface $\Sigma \sim \Lk \delta L_{crack}$, and we estimate the increment of $U_T$ around the crack as:
\begin{equation}
d U_T \sim - D \kappa^2 
L_k \delta L_{crack}. 
\end{equation}
The fracture threshold is reached when the energy release rate $G$ becomes equal to the surface energy $G_c$ ~\cite{tada2000, anderson2005, zehnder2012}. We then define the break-up criterion as: 
\begin{equation}
	G = G_c \sim D \kappa_c^2 \frac{L_k}{2h}
\end{equation}
We deduce that at the break-up threshold: 
\begin{equation}
    \kappa_c = \sqrt{\frac{2 G_c h}{D}} \Lk^{-1/2} = \beta \lambda^{-1/2}  ,
    \label{eq:Gc}
\end{equation}
where $G_c$ is the surface energy of the material, which is independent of the wave properties. 
We observe that $G$ is independent of $L_{crack}$, meaning that the crack propagation is marginally stable.
From the result of [Fig.~\ref{fig:4}c)], we deduce that the critical curvature for breakup $\kappa_c$ scales as $\kappa_c =  \beta \lambda^{-1/2}$ with $\beta = \sqrt{\frac{2 \alpha G_c h}{D}} = 5.82 \pm 0.6$ m$^{-1/2}$ (eq. \ref{eq:Gc}) ($0.6$ m$^{-1/2}$ being the standard deviation error from the fit). The uncertainty on the constant $\beta$ can be attributed to the variations of the material defects from one run to another, due to the varnish heterogeneity. The variations of $\beta$ with the varnish thickness is not significant enough to be conclusive.


\textit{Discussion} - In this study, we manage to form a thin crust of a brittle material floating on a water surface, and we measure its elastic properties using a non invasive technique. When excited by gravity waves, the crust breaks from bending and we show that the breakup threshold depends on the forcing geometry. The corresponding critical stress for fracture propagation also depends on the forcing wavelength. Our observations are incompatible with a critical strain criterion for fracture. A geometry dependent fracture threshold has been observed recently in a different configuration by Prabowo \textit{et al.} \cite{Prabowo2024}, in which they highlighted the role of the size of a spherical indenter on the fracture threshold.
Our fracture threshold can be rationalized using an energy criterion, which depends on the critical curvature $\kappa_c$, and the characteristic length on which the elastic energy can be released, $\Lk$. The threshold is controlled by a crack propagation criterion rather than a crack initiation criterion, as a consequence of the varnish heterogeneities from the grain scale to the sheet scale. The magnitude and the scale of these heterogeneities are most likely a significant source of errors and could explained the scatter of the critical curvature. However, since these variations are uncorrelated with the forcing scale $\lambda$, our main result, a fracture criteria that depends on the wavelength, still holds. 

\bapt{The break-up of sea ice by waves is also triggered by flexural stresses \cite{Squire1995}. Extrapolating our criterion to sea ice, the critical stress may depend on the forcing geometry, hence increasing with the wavenumber $k$, compatible with field observations~\cite{Kohout2016}. However, our experiments are conducted with non-linear and mainly mono-chromatic waves which is rarely the case in MIZ. For linear waves, we expect the same break-up criterion to hold with $\alpha = 1/3$. For a broader wave spectrum, to the best of our knowledge, there is no test of an energy criterion against field data.}

Currently, the \bapt{field} observations remain too scarce to isolate the effects of the local geometry on the fracture threshold \bapt{and cannot fully discriminate between break-up criteria}. We expect that further experiments on ice at various scales could help to better understand the fracture of sea ice by ocean swell.

 

    
\acknowledgments{We thank Ana\"is Abramian, Michael Berhanu, V\'eronique Dansereau, S\'ebastien Kuchly, Gabriel Le Doudic, Nicolas Mokus, Sylvain Patinet, Benoit Roman and Vasco Zanchi for fruitful discussions. We thank Amaury Fourgeaud and Laurent Quartier for technical assistance. This work has benefited from the financial support of Mairie de Paris through Emergence(s) grant 2021-DAE-100 245973, and from the Agence Nationale de la Recherche through grant MSIM ANR-23-CE01-0020-02.}

\bibliography{Baptbib.bib}

\end{document}